# Quantitative imaging of the complexity in liquid bubbles' evolution reveals the dynamics of film retraction

Biagio Mandracchia[1,*], Zhe Wang[1,2], Vincenzo Ferraro[3], Massimiliano Maria Villone[3], Ernesto Di Maio[3], Pier Luca Maffettone[3], Pietro Ferraro[1]

[1]CNR-ISASI, Istituto di Scienze Applicate e Sistemi Intelligenti «E. Caianiello» del CNR, Via Campi Flegrei 34, 80078 Pozzuoli, Napoli, Italy.
[2]College of Applied Sciences, Beijing University of Technology, 100124 Beijing, China.
[3]Dipartimento di Ingegneria Chimica, dei Materiali e della Produzione Industriale, Università di Napoli Federico II, Piazzale Tecchio 80, 80125 Napoli, Italy.

The dynamics and stability of thin liquid films have fascinated scientists over many decades. Thin film flows are central to numerous areas of engineering, geophysics, and biophysics and occur over a wide range of length, velocity, and liquid properties scales. In spite of many significant developments in this area, we still lack appropriate quantitative experimental tools with the spatial and temporal resolution necessary for a comprehensive study of film evolution. We propose tackling this problem with a holographic technique that combines quantitative phase imaging with a custom setup designed to form and manipulate bubbles. The results, gathered on a model aqueous polymeric solution, provide an unparalleled insight into bubble dynamics through the combination of full-field thickness estimation, three-dimensional imaging, and fast acquisition time. The unprecedented level of detail offered by the proposed methodology will promote a deeper understanding of the underlying physics of thin film dynamics.

## INTRODUCTION

Thin liquid films, such as soap bubbles, are ubiquitous in nature and technology. Biological vesicles, magma bubbles, insulating and food foams, detergents and oil foams, all share most of the physics, chemistry and engineering beneath the bubble formation and evolution[1,2]. Their study is also important since they mediate a wide range of transport processes encompassing applications from nanotechnology to biology[3–5]. These films may display unusual dynamics featuring the formation of regular or chaotic structures, periodic waves, shocks and fronts, and "fingering" phenomena.[6] The entire area is currently thriving with new discoveries, applications and, in particular, calls upon techniques to measure both the long-range thickness mapping and its fast acquisition on evolving thin-films. In fact, measurement of the thin film thickness evolution as a consequence of manipulation, drainage and rupture is key to understand such behaviors[7–9].

Nowadays, different techniques for Quantitative Phase Imaging (QPI) are used to measure the thickness of transparent three-dimensional objects with one dimension thinner than the other two (films)[10,11]. In particular, interferometry is routinely used for the study of thin fluid films and surface topology, using both monochromatic and white light[12,13]. Interferometry measures the intensity of fringes produced by the interference of light reflected at the two interfaces of a thin film. Such intensity depends on the wavelength of light, the refractive index of the sample, and the thickness of the material. These techniques can be divided in two families characterized by point-like or full-field inspection[9]. The first ones measure the thickness in a very restricted area of the film's surface. Early studies used a photomultiplier to precisely measure the equilibrium thickness of soap films contained in a special cell designed to isolate a thin film of liquid.[14] Modernized versions of this setup are currently used by diverse research groups.[15,16] Conversely, the full-field techniques measure the thickness across the whole surface of the film during the entire experiment.[17,18] Even though these systems can determine the film thickness with a resolution of few nanometers, they lack the (lateral) spatial or temporal resolution necessary to follow the complex dynamics of an evolving thin liquid film.

Here, we propose the adoption of a setup for the study of thin film dynamics based on off-axis Digital Holography (DH). Holographic microscopes are interferometers that allow for pseudo-3D reconstruction of objects captured out of the best focal plane. This feature adds flexibility to the experimental procedure and in turn, has kindled the spreading of DH beyond the field of metrology, from non-destructive testing for industry to label-free imaging of biological samples[19–22]. DH can accurately determine phase and amplitude by means of dense carrier fringes down to fractions of the illumination wavelength. A benefit of digital holography with carrier fringes is that unlike some other QPI techniques, e.g. phase shift interferometry[23], the necessary information is all gathered in a single frame, which is appropriate for high-speed data acquisition.

We report the measurement of thickness distribution all over an aqueous polymeric solution thin-film, during the formation of a bubble in non-ideal conditions, where a wide range of film thicknesses is present at same time in the film. Based on these data, the variation range and variation trend of the film thickness map are accurately measured, from the formation, to the inflation and the bubble rupture. In particular, during the bubble growth, the location of the bubble surface changes continuously so that an imaging systems in which the focusing of the image can be retrieved *ex-post* from the experimental recordings is required. DH allows such refocus of the sample by numerical processing of the recorded holograms[24]. In this way, it is possible to follow *a posteriori* the position of the film surface during the bubble formation.

## RESULTS

### Holographic thickness mapping for liquid films

The experimental setup was designed embedding a custom setup to form and manipulate a thin liquid film within an off-axis Mach-Zehnder interferometer, see Figure S1a. The films are formed on top of a metal pipe with an internal diameter of 18mm and a side inlet connected to a syringe pump, see Figure S1b-c. As a model system, we studied the temporal evolution of the thickness profile of bubbles formed from a film made of an aqueous solution of maple syrup and 0.05 wt% of polyacrylamide (PA). The bubbles were inflated by pumping air from the side inlet of the pipe at a flow rate φ = 0.015 mL/s[25]. DH in off-axis geometry is based on the classic holography principle, with the difference being that the hologram recording is performed by a digital camera and transmitted to a computer, and the subsequent reconstruction of the holographic image is carried out numerically, see Figure S1d-g. In DH, the interferometric acquisition system can only measure the phase modulo-$2\pi$, commonly referred as wrapped phase. To recover the absolute phase, and then thickness profile, we used the Phase Unwrapping Max-flow/min-cut (PUMA) method[26]. The PUMA method provides an exact energy minimization algorithm given the assumption that the difference between adjacent pixels is smaller than $\pi$ rad. From the experimental point of view, this leads us to make sure that we have a good sampling of the observed area in order to assume that thickness changes are sufficiently smooth in comparison to the fringe sampling, and no phase jumps are missed.

Once retrieved, the absolute phase gives a measure of the optical path length experienced by the laser beam, which is equal to the thickness of the film multiplied by its refractive index. Thus, knowing the refractive index of the solution bunches used in the experiments (see Supplementary Information), we can easily map the evolution of the film thickness during the bubble growth and drainage, see Figure 1a and Supplementary Video 1 & 2.

Finally, DH acquisitions are pseudo-3D representations of the optical thickness of the sample. This means that the measured thickness profile, $s$, is a projection on the image plane of the three-dimensional structure of the sample, see Figure S3. However, the thickness along the normal to the bubble surface, $\bar{s}$, can be retrieved by geometrical considerations, see Figure 1b-d and Supplementary Video 3. It is worth noting that in correspondence of the center the two values are almost identical. For example, within 1.3mm from the center, the estimated relative error is less than 1%, see Figure S4.

### Film thinning and bubble growth

The shape of the bubble in our system is mainly controlled by the volumetric air flow, $\varphi$, set by the pump, and, if $\varphi$ is constant, the volume of the bubble grows linearly in time:

$V_{bubble} = \varphi t$   (1)

Considering the bubble as a spherical cap of height $h$ and basal radius $a$, we can rewrite the previous equation as:

$\frac{\pi}{6} h(3a^2 + h^2) = \varphi t$  (2)

The geometric parameters of the bubble, then, can be fully controlled by the pump.

To study the film thinning due to the gravitational drainage of the fluid along the bubble surface, we adjusted the experimental parameters in order to maximize the bubble stability while approaching the hemispherical shape. We observed that reasonably stable bubbles could be formed inflating air into the metal pipe with a relatively low flow of φ=0.015mL/s. Nonetheless, we found it difficult to reach a perfect hemispherical shape of the bubble (h~a). Furthermore, this configuration resulted impractical for the study of drainage towards the borders, as discussed in the previous section, therefore we decided to stop the pump at a height of approximately two thirds of the basal radius (h~2/3a).

Bubbles were observed from the top and from the side. The top view is recorded by a CCD camera at a maximum frame rate of 60Hz. The side view is recorded by a CMOS camera (Apple Inc. iSight) at 30Hz, see Supplementary Video 4. The experiments were conducted at 23°C.

During inflation, $h$ is a function of time and equation (2) can be rewritten as:

$h(t)(h^2(t) + 3a^2) = \frac{6\varphi}{\pi} t$ (3)

From which we can derive the formula:

$h(t) = \frac{\sqrt[3]{\sqrt{4a^6 + b^2 t} + bt}}{\sqrt[3]{2}} - \frac{\sqrt[3]{2} a^2}{\sqrt[3]{\sqrt{4a^6 + b^2 t} + bt}}$   (4)

with $b = \frac{6\varphi}{\pi}$.

A good fit of the experimental data is given by a first order approximation of Equation (4), see Figure S1h:

$h(t) = \gamma \sqrt[3]{1 + \beta t} - \frac{\gamma}{\sqrt[3]{1+\beta t}}$   (5)

The thickness maps in Figure 1e show an accumulation of the fluid in the central region, when the bubble is still flat (prior to pump starting, $t$=0.1s). During the inflation, it is possible to observe a gradual thinning at the center that slowly continues when pumping is stopped. This process is the consequence of the gravitational drainage of fluid from the top towards the rim of the bubble.

Gravitational drainage causes the film thickness to decay exponentially with time [27]. Now, recalling that in the center $s = \bar{s}$, we have that:

$s = s_0 e^{-\frac{t}{\tau}}$   (6)

where

$$\tau = \frac{\alpha}{h} \; ; \; \alpha = \frac{\mu}{\rho g}. \quad (7)$$

with $\mu$ being the viscosity, $\rho$ the liquid density, $g$ the gravitational acceleration, and $s_0$ the initial thickness, see Figure 1f.

During inflation, the film thickness of the bubble has a more complicated dependence with time. Indeed, the drainage is concurrent with the film stretching as a consequence of the increase of the bubble surface. However, the experimental data can be satisfactorily approximated also by a linear function (see Figure 1g):

$$s = s_0(1 - \beta t). \quad (8)$$

**Fluid drainage and convection**

The continuous drainage towards the borders causes a decrease of the mass of the fluid with time. This is directly proportional to the volume of the film layer: $V = M/\rho$. Ideally, if the fluids were perfectly homogeneous, we would expect the drainage to be radial. This means that the thickness of the film does not depend on the polar angle but only on the latitude. This assumption fails for real films, where some level of inhomogeneity or asymmetry is present in the system and gives rise to phenomena such as fluid's convection inside the film. As expected, the center of the bubble tends to get thinner, the bigger the bubble (Figure 2a-b and Supplementary Video 5). Nonetheless, this phenomenon is not homogenous. At the same time, it can be noted that this change in thickness does not happen uniformly, but it seems to be related to a momentary rearrangement of the fluid across the surface.

After an initial stasis period, Figure 2c (gray area), the bubble volume drops with time and follows the expected exponential decay (red dashed line). Surprisingly, after reaching a plateau value, the volume starts to grow again a few seconds before the rupture, see Figure 2c (yellow area). A more detailed analysis reveals that this increase can be related to a change of the drainage dynamics of the fluid, see Figure 2d-e. It is possible to devise two different contributions to this inversion of trend, one at the center of the bubble and the other close to the edge of the pipe. The first contribution is due to a relatively small increase of the thickness of the film around the center, Figure 2d-f (red arrows). The second contribution is given by a steady in-flow of part of the fluid from the edge of the pipe back towards the center of the bubble, Figure 2d-f (black arrows). This in-flow takes the form of a regular pattern, which can be devised after 4.5s. The regular patterns observed at the latest stage of the film evolution dynamics, depicted in Figure 2b, have been observed elsewhere, for vertical and horizontal thin films, and usually addressed to as "fingers" and due to Marangoni effect, Plateau-Rayleigh instability and/or marginal regeneration. In the context of Plateau-Rayleigh instability, in a cylindrical flow with infinite length, the characteristic wavelength of the pattern is [28]:

$$L = \frac{s}{4\pi}\sqrt{2 + 3\sqrt{2}Oh}, \quad (9)$$

where $Oh = \frac{\mu}{\sqrt{2\gamma\rho s}}$. In our case we found $L$=2.2mm, which is of the same order of magnitude of the experimental value $L_{exp} = 2\pi R/N$=5.4mm, where $N$ is the number of fingers on the image ($N$=9) and $\mu$ is the viscosity. Furthermore, the motion of these patterns could be due to the marginal regeneration: along the edge of the film, where the film connects with the pipe, there is a "Plateau border" that has got curved surfaces and lower Laplace pressure than the central part of the film; thicker parts are drawn bodily into the border by the negative excess pressure while the thinner film is pulled out of the border.

In order to give a plausible physical interpretation to the experimentally-observed non-monotonic trend of bubble thickness shown in Figure 2c, we have performed a Finite Element numerical simulation of a system mimicking the experimental one. The mathematical model underlying the numerical simulation is constituted by the mass and momentum balance equations and the constitutive equation for the liquid film supplied with proper boundary and initial conditions and it is detailed in the Supplementary Information. The constitutive parameters of the liquid have been derived from the rheological data of the fluid employed in the experiments, see Supplementary Information.

The numerical temporal trend of the thickness at the center of the film h, normalized by its initial value $h_0$, is reported in Figure S2. By comparing Figures S2c and 2c, it is apparent that, even if in the simulation a simplified system is considered, a good agreement holds between the experimental behavior of the volume of the bubble central portion and the simulated evolution of the film thickness, with an initial steep decrease while the bubble is inflated, an almost horizontal portion, and, then, an inversion of the trend, i.e., a thickening at the center of the bubble.

From the outcome of the numerical simulation, the latter can be ascribed to a fluid drainage from the rim deposited on the pipe edge toward the center of the film due to surface tension. Indeed, two opposite mechanisms act: during inflation, the film thins at the center due to gravity and liquid adherence at the pipe wall, whereas surface tension makes the fluid go from the border to the center to minimize the film external surface. Since inflation is fast, at the beginning the effects connected to it dominate, then, when inflation ends, the "reservoir" constituted by the rim "pumps" the liquid back, thus making the film thicken at the center.

Figure 3a depicts a thin liquid film thickness evolution in a slightly different case in which the film is left for a long period of stasis before the inflation. In fact, thickening in the central part of the film is observed, due to sagging. Moreover, the topography of the film appears less homogenous than in the case of Figure 2, as a further evidence of the thickness measurement accuracy as by the proposed technique, particularly when inhomogeneities are present. In these cases, the mass of fluid tends to accumulate at the center of the film before the inflation begins, see Figure 3a-b. However, it quickly drains towards the rim once the bubble starts growing. In this case, the dynamics of the drainage process are not only far from ideal but, at certain moments, the whole process seems to stop, see Figure 3c (yellow areas). On the other hand, the rate of volume drainage does not go to zero, nor in time nor along the radius, see Figure 2d-f. This supports the fact that the rapid movement of a big quantity of mass creates some complex movement of the fluid at the rim, which could temporarily counterbalance the draining process, see Supplementary Video 6.

**Flow tracking**

It is generally hard to describe a situation of complex motion such as the one depicted in Figure 3. On the one hand, we have shown how it is possible to estimate the drift of the fluid and the dynamics of formation and dissolution of mass aggregates due to the presence of fluid vortexes. On the other hand, the assessment of the dynamics of liquid film rearrangement can be simplified following the displacement of particles dispersed in the fluid by holographic three-dimensional tracking.

We injected PMMA particles with nominal diameter of 6μm into the PA solution and tracked them in three dimensions by automated numerical refocusing, Figure 4. After holographic amplitude reconstruction, three particles were selected from all visible points, which follow different paths along the bubble surface. To effectively identify and assess the movement of these particles, we used the correlation recognition tracking method[29,30].

All three particles have different trajectories and show non-trivial flows within the film. Indeed, it is possible to observe how they can have both radial and swirling motion. The observed speed of the process and its span in the third dimension make the tracking of the particles hard with standard imaging techniques. Holographic 3D tracking, conversely, has proven suitable for these situations. This piece of information can be useful to analyze the mass flow on bubble film surface but also to follow the arrangement of colloids inside the film.

**High-speed holographic imaging**

The rupture of a bubble is a very fast process that requires the use of high-speed cameras to be observed. It can have very different dynamics, depending on the particular fluid or conditions of breakage[31]. One important parameter is the thickness of the opening rim and the possible presence of fluid droplets escaping from the film[25,32].

To induce the rupture, we placed a needle on top of the metal pipe we use to grow the bubble. When the bubble reached an almost hemispherical shape, we gently lowered the needle until it was in contact with the film. To record the bubble rupture, we used a high-speed CMOS camera (Mikrotron, MC1310, 980 Hz).

Before the bubble rupture, it is possible to see a black film forming in correspondence of the tip of the needle, Figure 5a-c (white dashed line). The black film forms where the film thickness is half the illumination wavelength, when the local destructive interference cancels the light passing through. In holographic reconstructions, this local absence of light is associated to the generation of random values. This is why, in phase images, black spots correspond to areas of low SNR. Finally, after around 453ms, the boundary breaks and the bubble opens, Supplementary Video 7.

High-speed holographic imaging can be useful to study the mechanics of bubble rupture in deeper detail. Quantitative thickness mapping is essential to distinguish the diverse profiles of the hole's rim, which characterize the retraction behavior of fluids[33]. Moreover, when asymmetric breakage profiles are observed[34], thickness mapping provides a link between the rupture path and the topography of the film.

Here, we observe for the first time the mass accumulation at the rupture edge during the film retraction, in accordance to the model proposed in Ref. 33, see Figure 5d. During retraction, the film tends to accumulate at the rim and then it becomes flatter during the last moments of the breakage. Also, the thinning process steadily continues and the black film rapidly expands around needle. However, further analysis reveals that the boundary seems to move faster along the directions where the film is thinner ($\theta = 180°$ and $270°$), probably following a least-resistance path, see Figure S3.

## DISCUSSION

The study of thin films and bubble rupture is of great interest to industrial processes and life science. Indeed, foams as well as plasma membranes or vesicles can be modeled in a manner similar to soap films and bubbles. The nature and properties of such structures have been the subject of extensive studies and continue to be attentively investigated[35].

The characteristics of these systems representing the hardest characterization challenges can be summarized as follows. First of all, they have fast and ever-changing dynamics. Real-time imaging systems and possibly fast recording devices should be used. Secondly, the film thickness varies from tens of micrometers to few hundreds of nanometers. This depends strongly on the nature of the solution and on the experimental conditions utilized for film formation. Finally, yet importantly, the bubble film is not uniform. This means that the bubble surface is a complex system and it will have a unique structure each time a new sample is prepared. The distribution of the polymer across the film changes every time and, even under the same pumping conditions, the time to rupture is not barely constant. Using air flow at 0.015 mL/s, we saw this time going from 3s to 10s. It is likely that such a difference is due to both the initial bubble thickness and the evolution dynamics. This is why evaluations of the film thickness based on geometrical considerations are not sufficient; instead, a continuous and quantitative inspection is necessary.

In this work, the design and implementation of a setup for the imaging of the dynamics of thin bubbles is presented. Our setup is based on DH to obtain quantitative images of the sample film dynamics. Along the years, many methods based on interferometry have been proposed to measure the actual film thickness and to monitor the interfacial rheological properties of these systems. Differential interferometry methods have also been described in investigations of contact angles[36] and bubble caps[37]. A method based on phase shift interferometry was developed for measurements on vertical films[38]. Other approaches based on resonant differential interferometry, fringe patterns from a dual-wavelength reflection, and speckle interferometry were also reported[39,40].

Compared with the abovementioned interferometric techniques, DH combines several advantages. Fist, the spatial resolution is limited only by the optics used and this is not true for methods that use color matching, where the thickness is measured in few points and then interpolated over all the image [41]. Second, it gives the full-field three dimensional information of the sample, unlike techniques that use photomultipliers to have very fast measurement but only in one point [14]. Third, DH does not need multiple exposures and can be matched with high-speed cameras to measure rapidly changing features[23], such as the rim of the hole formed by the rupture of the liquid film.

Owing to the spatial resolution and fast, full field measure of the liquid film thickness, we proved that this technique has several novel features. In Figure 3 we showed the time dependence of the film thickness on an evolving geometry of the bubble (in the past, the only way to measure bubble's thickness was without or after inflation). In Figures 4 and 5 we showed the film volumes evolution and, for example, observed that, in the last few seconds, there is a thickening of the average film thickness, although drainage towards the bottom would have suggested a monotonic reduction of the average thickness. This thickening is caused by mass fluxes from the border of the film through the center as it is shown in Figure 2 using gradients plots. As such, this technique can be used to investigate phenomena like tear spreading or coffee rings formation,[3,42,43] where Marangoni effects, drainage and wetting concur to the thin film evolution.

In conclusion, we proposed an experimental setup that gathers for the first time all the features required to study the liquid thin film evolution. Nevertheless, this comes at the price of a more complex data analysis. However, there are now diverse resources available for both hologram reconstruction and data analysis, so that a custom code is seldom needed. Furthermore, the local thickness is calculated assuming a certain degree of continuity, i.e. step heights of less than half a wavelength. Even if the results do not show any contrary evidence and seem in agreement with the expected values, it can still be viewed as a limitation. This limitation can be overcome by changing the system used for the bubble formation, e.g. growing bubbles on top of a glass surface. Then, again, future work should focus on the implementation of a robust optical solution, such as dual wavelength DH. This method can considerably extend the dynamic range of phase detection, removing most of the issues related to phase wrapping.

The application of DH is not limited to the proposed configuration but could be adapted without difficulty to different ones. The metal pipe, for example, can be replaced by a quartz cuvette, which would be useful to study the formation of gas bubbles in a fluid[41], Figure S4a. Alternatively, a configuration similar the one proposed for phase shift interferometry can be used to study spherical bubbles pending from a nozzle[23], Figure S4b. If controlling the volume is not essential, the bubbles could be grown on a glass substrate or a petri dish[27], as in Figure S4c. In this case, the illuminating beam could be slightly tilted in order to avoid illuminating the needle. In this way, it is possible to image the very first moments of the bubble rupture close to the tip of the needle. Finally, virtually all the systems currently used for the study of flat bubbles could be easily integrated into a holographic microscope[7,8,18], Figure S4d.

## MATERIALS AND METHODS

### Experimental setup

The DH setup consisted of an off-axis Mach-Zehnder interferometer with a sample stage adapted for the control of bubble growth. The experimental setup is schematically shown in Figure S1a. The illumination source was a HeNe laser ($\lambda$ = 632.8nm). In the Mach-Zehnder interferometer, the laser beam is divided in two parts by a polarizing beam-splitter cube (PBS). The resulting beams are referred to as object and reference beam. The object beam illuminates the sample from the top and forms the image on the camera. On the contrary, the reference beam goes towards the camera without passing through the sample. The two beams are collected by a second beam-splitter cube (BS), which is slightly tilted so that the two beams overlap with a small angle. This angle controls the period of the interference fringes and can be adjusted according to the sampling requirements. The image of the sample is de-magnified by a factor 0.25 with two lenses put in front of the camera (f= 200mm and 50mm, respectively). With an estimated maximum diameter of the circle of confusion of 0.4mm, the depth of focus of the system is 8mm.

### Bubble formation

The bubble growth was controlled using a custom metal pipe, see Figure S1b-c. The pipe had a diameter of 18mm and a side inlet that was connected to a syringe pump (Harvard Apparatus). The rim of the top of the pipe was slightly grooved in order to maximize the contact surface with the bottom of the bubble. An aqueous solution of maple syrup (Maple Joe, Famille Michaud Apiculteur, Gan, France) and 0.05 wt% of polyacrylamide (Saparan MG 500, The Dow Chemical Company, Midland, MI, USA) was used. Bubbles were created forming a film made of the solution on top of the pipe and placing the bottom on a glass Petri dish. The pipe was secured to the sample holder to prevent any possible movement during the measurements. Finally, a syringe pump (Harvard Apparatus, Model 22) was utilized to inflate the film and form a half bubble with a flow rate of 0.015 mL/s. A small amount of water was added on to the dish to avoid pumped-air leakage.

### Wavefront reconstruction

Digital holography in off-axis geometry is based on the classic holography principle, with the difference being that the hologram recording is performed by a digital camera and transmitted to a computer, and the subsequent reconstruction of the holographic image is carried out numerically.

The recorded intensity $I_H(x_H, y_H)$ at the hologram plane is the square module of the amplitude superposition of the object and reference waves. It is given by:

$$I_H(x_H, y_H) = |O_0(x_H, y_H)|^2 + |R_0|^2 + O_0^*(x_H, y_H)R_0 + O_0(x_H, y_H)R_0^*$$

The phase information of the hologram is provided only by the last two terms, which are filtered and centered in the Fourier space. We reconstructed the holograms by numerically propagating the optical field along the z direction using the angular spectrum method. If *E(x,y;0)* is the wavefront at plane *z=0*, the angular spectrum *A(ξ,η;0)=F{E(x,y;0)}* at this plane is obtained by taking the Fourier transform, where *F{}* denotes the Fourier transform; ξ and η are the corresponding spatial frequencies of x and y directions, respectively; and *z* is the propagation direction of the object wave. The new angular spectrum A at plane *z = d* is calculated from *A(ξ,η;0)* as:

$$A(\xi, \eta; d) = A(\xi, \eta; 0) \cdot \exp\{j \frac{2\pi d}{\lambda} [1 - (\lambda\xi)^2 - (\lambda\eta)^2]^{\frac{1}{2}}\}$$

The reconstructed complex wavefront at plane *z=d* is found by taking the inverse Fourier transform as

$$E(x, y; d) = F^{-1}\{A(\xi, \eta; d)\}$$

where *F⁻¹{}* denotes the inverse Fourier transform. The intensity image *I(x,y;d)* and phase image *φ(x,y;d)* are simultaneously obtained from a single digital hologram by calculating the square module of the amplitude and the argument of the reconstructed complex wavefront:

$$I(x, y; d) = |E(x, y; d)|^2$$

$$\phi(x, y; d) = \arctan\left(\frac{Im[E(x, y; d)]}{Re[E(x, y; d)]}\right)$$

The workflow of numerical reconstruction is shown in Figure S1d.

From the experimental data, we observed that for each frame's spectrum, the +1 order center changes with the bubble growth. We supposed that this phenomenon was due to the bubble surface change. The bubble surface could be seen as a lens, twisting the object beam and slightly changing the off-axis angle during the growth. Therefore, if we use the same filtering window for each frames of holographic video, in the final phase result, we would get a random phase distortion, and it would greatly affect the phase measurement accuracy. This issue was addressed using an automatic filtering algorithm during the holographic video reconstruction. This algorithm simply scans the Fourier spectrum for the maximum of the +1 diffraction order and centers the filter accordingly.

## Thickness estimation

In DH, thickness estimation is directly related to the accuracy of the absolute phase recovery. Indeed, an interferometric acquisition system can only measure the phase modulo-$2\pi$, commonly referred to as wrapped phase. Formally, we have $\phi(x, y; d) = \psi(x, y; d) + 2k\pi$. Where $\phi$ is the absolute phase value, $\psi$ is the wrapped phase, i.e. the measured value, and $k \in \mathbb{Z}$ is an integer accounting for the number of $2\pi$ multiples.

The main task of a phase unwrapping algorithm is the choice of the place where the phase of the field should be shifted. In real experiment conditions such choice is often complicated by a phase noise which can lead to erroneous phase unwrapping shifts. Since phase noise often have a higher frequency than the desired signal, initial filtering of the wrapped phase field is the easiest and most intuitive way to simplify unwrapping[44].

In this work, we used the Phase Unwrapping Max-flow/min-cut (PUMA) method[26]. The PUMA method provides an exact energy minimization algorithm given the assumption that the difference between adjacent pixels is smaller than π rad. From the experimental point of view, this leads to assume that film thickness changes are smooth enough to be well sampled by the camera pixel. This assumption can be aided by changing the magnification of the system accordingly to the homogeneity of the sample. A good sampling of the observed area is very important for an accurate thickness estimation. Indeed, when peaks or valleys are too steep in comparison to the fringe sampling, some phase jumps can be missed and a wrong absolute phase recovered.

Furthermore, absolute phase estimation requires the assessment of a possible bias unwrapped and real phase profile. This is usually done taking a reference point within the field of view. Here, this reference point is given by the Newton black films that form during bubble inflation. The absolute thickness of these areas is half the illumination wavelength (in our case 316nm). In the frames where these are not present, we assumed that the process of film thinning is continuous and slow in comparison to the recording speed. At the end of the bubble's life, i.e. when this is close to rupture, this assumption is not necessary because black films are usually present. However, more complicated situations in which the estimation of a reference thickness may be harder can be addressed adopting one of the variant systems proposed in the Discussions section and sketched in Figure S3.

Once obtained the absolute phase map, the local thickness estimation is given by the formula: $s = \frac{\lambda}{2\pi}\frac{\phi}{n-1}$, where λ is the illumination wavelength and *n* the refractive index of the solution. In this estimation *n* is considered a constant and this could induce some error when this is not true. In liquid films, of course, water evaporation alters the density of the solution and, in turn, the refractive index *n*. However, at the time scale of our experiments, we estimated that the error related to evaporation is negligible ($\delta s < 5\%$), see Supplementary Information.

It is well known that holographic measurements yield pseudo-3D images. This means that the measured thickness profile, *s*, is a projection on the image plane of the three-dimensional one. However, the radial thickness, $\bar{s}$, i.e. the thickness along the normal to the bubble surface, can be retrieved by geometrical considerations, see Figure S6. Assuming that the upper and lower surface of the bubble are locally parallel, we have that $\bar{s} = s\sqrt{1 - \frac{r^2}{R^2}}$, where *r* the distance from the center in the image plane and *R* is the radius. It is worth noting that in correspondence of the center the two values are almost identical. For example, we estimated a relative error $\frac{\bar{s}-s}{s} = 1 - \sqrt{1 - \frac{r^2}{R^2}} < 1\%$, for $r < 1.3mm$, see Figure S7. On the other hand, in proximity of the pipe's border, the presence of meniscus deformation alters the estimation of $\bar{s}$, see Figure S8. For this reason, we used $s$, instead of $\bar{s}$, to calculate the volume or the draining rate at the borders because it conveys the same information with less geometrical assumptions, see Figure S9.

Herein, we used aqueous solutions that were homogenous enough not to need any particular adjustment of the optical setup. However, when this is not the case and particularly inhomogeneous samples are to be studied, the use of two or more beams with different wavelengths is suggested [45]. Using different illuminating wavelengths with close values gives the possibility to create a synthetic one with a bigger value, and therefore to enlarge continuous phase regions of the reconstructed wavefront. Often with this method the unwrapping procedure is simplified or not required at all. Nonetheless, holograms registration with different light wavelengths results in a more complicated technique both in hardware and software.

Supplementary information accompanies the manuscript.

## CONFLICT OF INTEREST

The authors declare no conflict of interest.

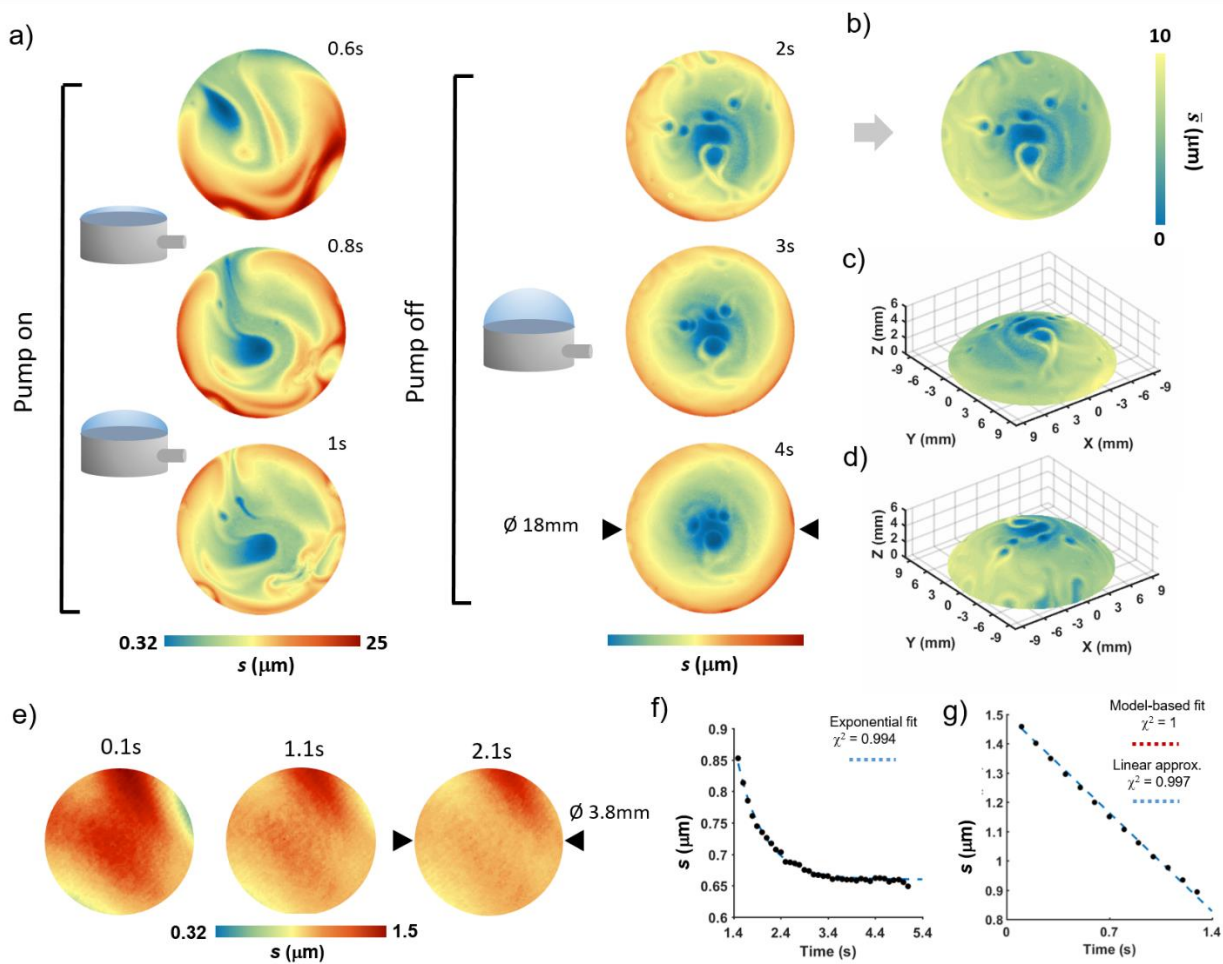

**Figure 1: Holographic thickness mapping during inflation and drainage.** (**a**) Evolution of the film thickness during bubble inflation (left) and drainage (right). The thickness values were obtained by holographic measurements, knowing that the refractive index of the sample was 1.47. During the experiment the film was inflated for 2 seconds and then let drain naturally until rupture. (**b**) Corrected map of the film thickness. Assuming that the bubble surface can be approximated by a spherical cap, it is possible to retrieve the film thickness in the radial direction. Three-dimensional depictions of the radial thickness map are shown in (**c**) and (**d**). (e) Drainage and film thinning at the center of the bubble. Thickness maps of the center of a bubble obtained by Digital Holography. The bubble was let grow for 14s. Afterwards, the pump was turned off and the fluid let drain naturally until rupture. Scale bar 1mm. (**f**) Plot of the thickness as a function of time during bubble blowing. (**g**) Plot of the thickness as a function of time during gravitational drainage.

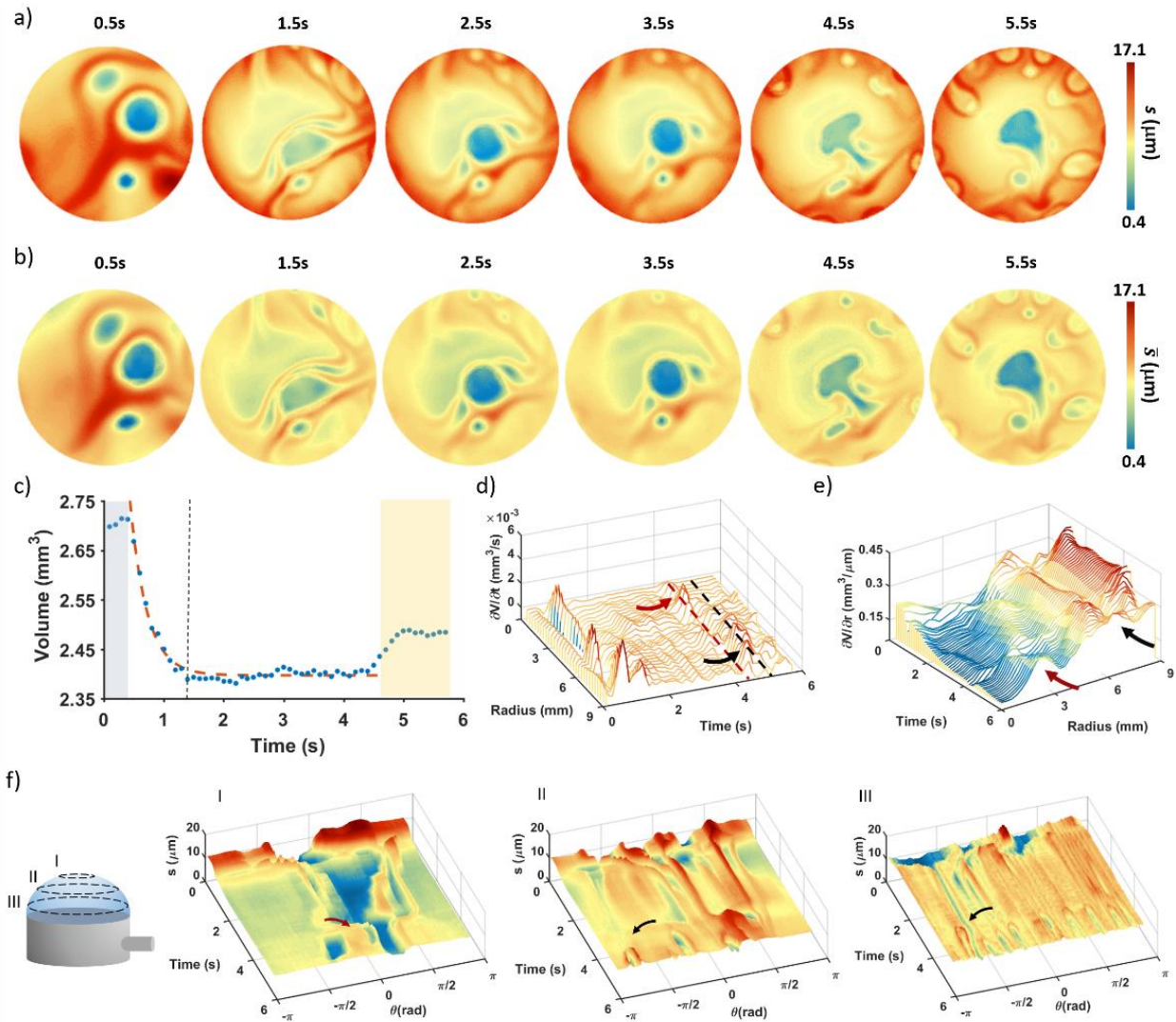

**Figure 2: Fluid drainage.** Thickness maps of the whole bubble surface before (**a**) and after (**b**) curvature correction. At first, the fluid accumulates in the center with no particular ordering. It is possible to see very thin areas in the film randomly positioned. While the bubble is growing, however, most part of the fluid is drained towards the edges and the thin films move to the center. (**c**) Plot of the Volume as a function of time. After an initial stasis time, the mass drainage appears to follow the expected exponential behavior. Just before the rupture time, on the other hand, the volume starts to increase once again. The derivatives along time (**d**) and radius (**e**) show that this volume increase goes from the rim toward the center (see black arrows). (**d**) Thickness change vs Time and azimuthal angle ($\theta$) at three different distances from the bubble center, namely I) 3mm, II) 6mm and III) 9mm. It is possible to observe some amounts of mass flowing back from the rim to the center.

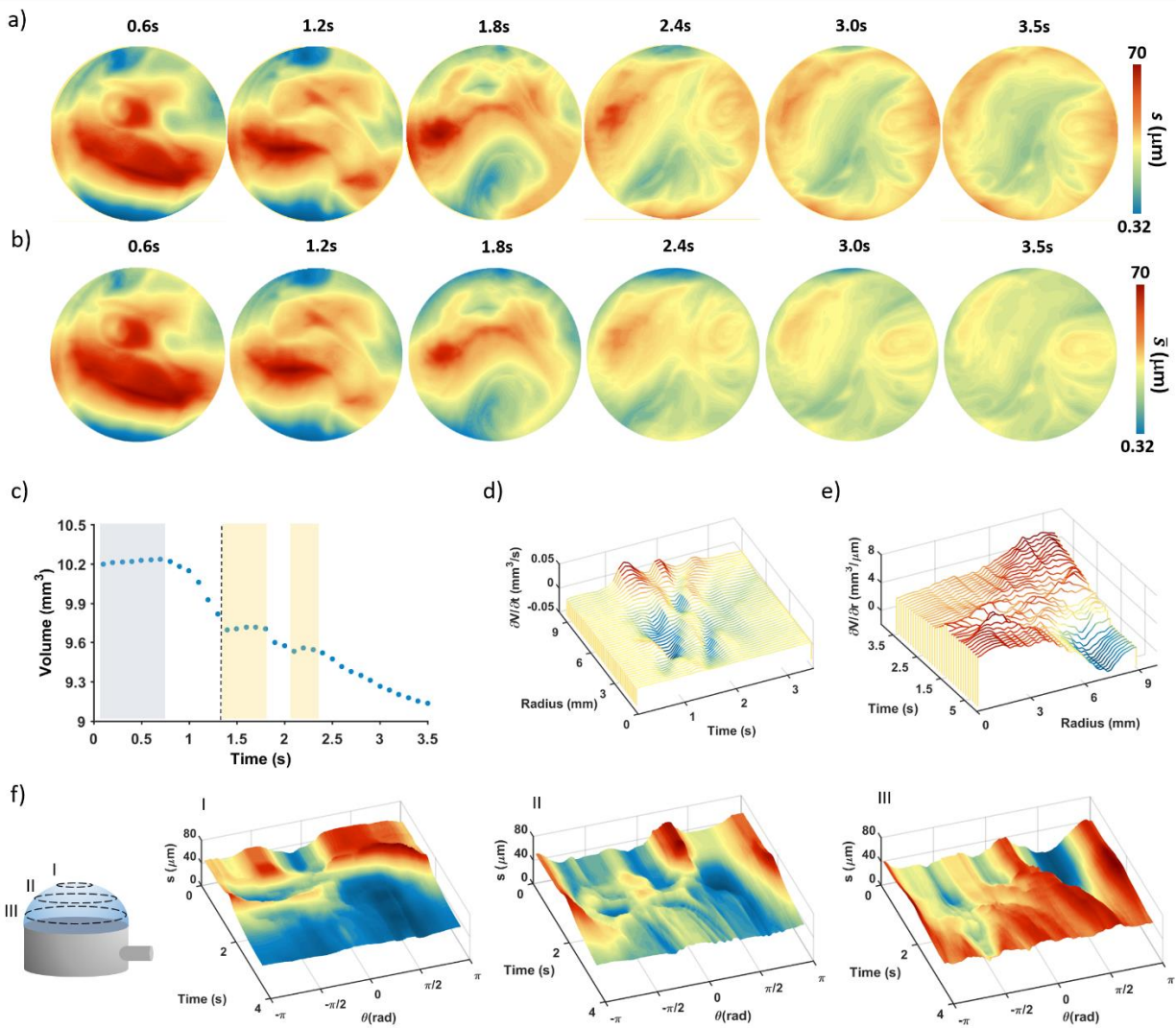

**Figure 3: Complex fluid drainage.** Thickness maps of the whole bubble surface before (**a**) and after (**b**) curvature correction. At first, the most part of the mass accumulates in the center. While the bubble is growing, instead, the fluid is drained towards the edges. This process is chaotic and includes the presence of vortexes. (**c**) Plot of the Volume as a function of time. The non-ideal nature of the drainage causes part of the fluid to come back towards the center of the bubble. This explains the presence of brief plateaus. Derivatives of the volume along time (**d**) and radius (**e**). (**f**) Thickness vs Time and azimuthal angle (θ) at three different distances from the bubble center, namely I) 3mm, II) 6mm and III) 9mm. The asymmetry of the process is especially clear in sections I and II.

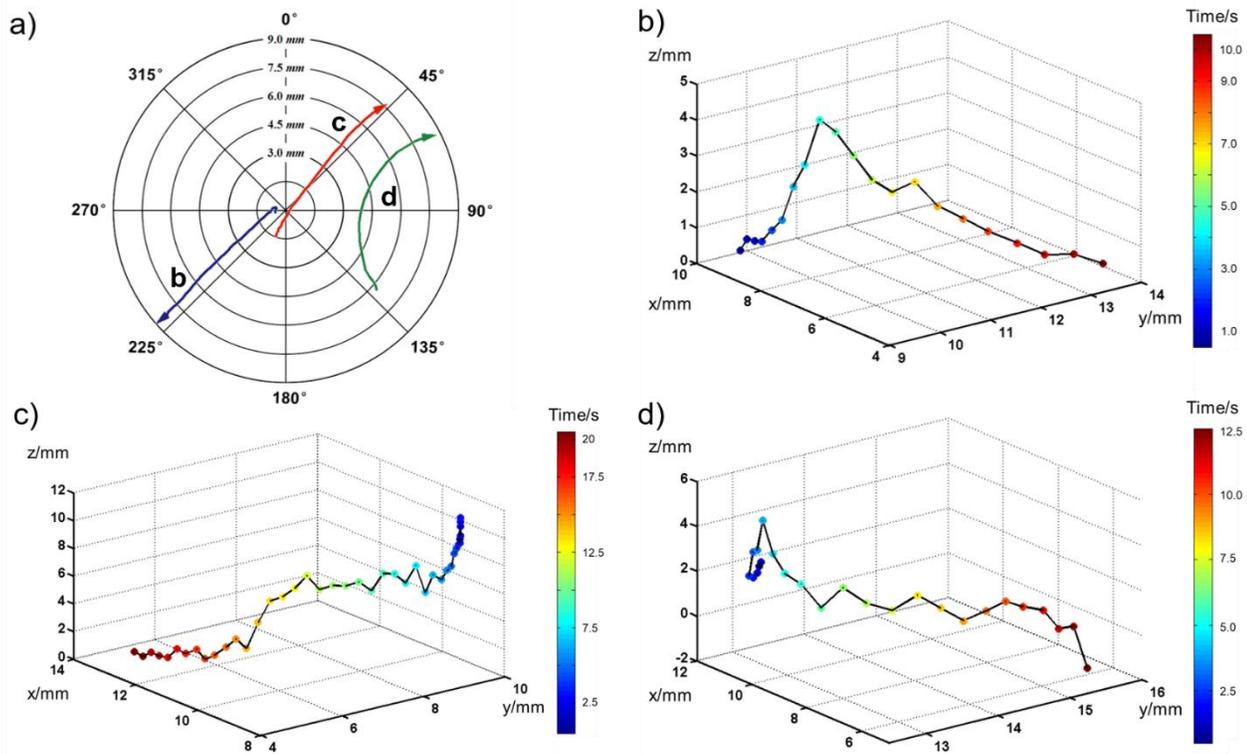

**Figure 4: 3D tracking of PMMA particles diffusing in a liquid film.** (**a**) Top view of the motion of three particles diffusing on the bubble surface. Their motion shows the presence of radial mass currents and vortexes. (**b-d**) three-dimensional plot of the motion of each particle.

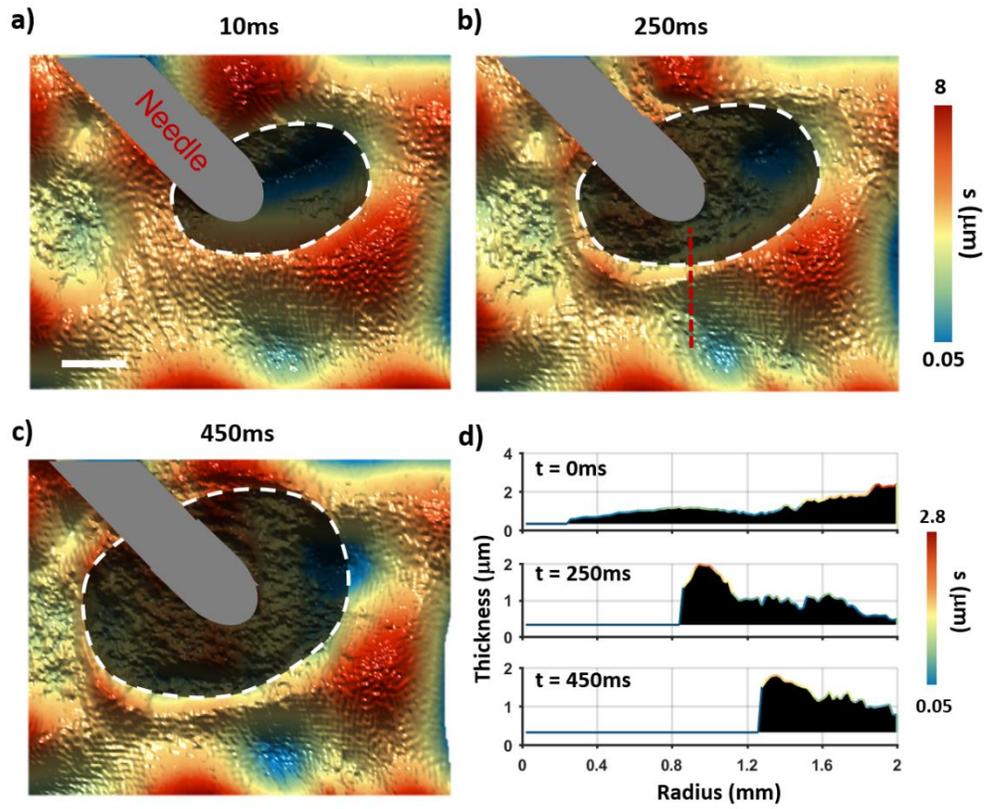

**Figure 5: High-speed holographic imaging of bubble rupture.** (**a-c**) Formation of the black spot around the needle tip (white dashed line). the black area indicates the phase noise related to the absence of light. The gray area indicates the position of the needle. Scale bar 1mm. (**d**) Thickness profile at the rim of the black film (red dashed line). During retraction, the film tends to accumulate at the rim until rupture.